\documentclass[english,aps,showpacs,superscriptaddress,preprintnumbers,amsmath,amssymb,nofootinbib,twocolumn]{revtex4}

\usepackage{graphicx}

\newcommand{\STr}{\mathrm{STr}}

\newcommand{\E}{\mathrm{e}}
\newcommand{\I}{\mathrm{i}}
\newcommand{\be}{\begin{eqnarray}}
\newcommand{\ee}{\end{eqnarray}}

\newcommand{\nn}{\nonumber }

\newcommand{\fderL}[1]{\frac{\overrightarrow{\delta}}{\delta #1}}
\newcommand{\fderR}[1]{\frac{\overleftarrow{\delta}}{\delta #1}}

\usepackage{babel}
\makeatother

\begin{document}

\title{Thermodynamics of QCD low-energy models\\ and the derivative expansion of the effective action}
\date{\today}                                           
\author{Jens~Braun}
\affiliation{Theoretisch-Physikalisches Institut, Friedrich-Schiller-Universit\"at Jena, 
Max-Wien-Platz 1, D-07743 Jena, Germany}
\affiliation{TRIUMF, 4004 Wesbrook Mall, Vancouver, BC V6T 2A3, Canada}

\begin{abstract}
We study the equation of state of a strongly interacting theory of relativistic bosons
and chiral fermions in the vicinity and above the chiral phase transition temperature.
Our model resembles presently used low-energy models of QCD in many ways, but
its simplicity allows us to study systematically various approximation schemes by means of a
derivative expansion. In particular, we compare the results for the phase transition temperature, 
equation of state and the thermal mass of the scalar fields obtained in various approximations.
We find that the large-$N_c$ approximation, being the zeroth order of our approximation, deviates significantly
from approximation schemes in which effects of (Pseudo-) Goldstone modes have been taken into account,
even at high temperatures and even if we allow for a finite explicit symmetry breaking. The important
role of the (Pseudo-) Goldstone modes is also manifested in the phenomenon of "local order" which 
is missing in large-$N_c$ studies.
\end{abstract}


\maketitle

\section{Introduction}
The thermodynamics of strongly interacting relativistic bosons and fermions play a prominent role 
in various fields of physics, such as heavy-ion collision experiments. There the theoretically
computed equation of state of QCD is used as an input in order to describe the expansion
of the so-called fireball~\cite{BraunMunzinger:2003zz}. However, the computation of the equation of
state is a highly-involved problem since the high-temperature limit is not a weak-coupling limit, at least 
for the soft modes, see e.~g.~\cite{Appelquist:1981vg,Kajantie:1995dw}. 
On the contrary, by increasing the temperature we interpolate continuously between the 
$4d$ theory and the underlying $3d$ theory which does not generally have to be weakly coupled. Indeed, 
QCD in $3d$ as well as simple scalar field theories in $3d$ have shown to be strongly coupled. 
For this reason, a naive perturbative expansion of the partition function at high temperatures is bound to fail. 

Several methods have been developed to study the equation of state. Lattice simulations are definitely the most 
powerful tool that we have currently at hand for a computation of the equation of state of strongly coupled 
theories. However, since lattice simulations are performed in a finite volume and at finite explicit symmetry breaking 
(i. e. current quark mass in the QCD context or external magnetic in the context of Ising models), the result for the 
equation of state has to be extrapolated to the continuum limit. The presence of dynamical fermions 
complicates lattice simulations further. On the other hand, continuum approaches have to deal with 
complementary problems, such as necessary truncations of the action of the theory.

One prominent approach to a systematic study of the thermodynamics of strongly interacting theories are 
hard-thermal loop resummation techniques, see 
e.~g.~Refs.~\cite{Pisarski:1988vb,Pisarski:1988vd,Frenkel:1989br,Braaten:1990az,Thoma:2000dc,Blaizot:2001nr,Blaizot:2003tw,DiRenzo:2006nh}.
For example, it has been shown in this framework that the momentum dependence of the $n$-point 
functions longitudinal and transversal to the heat-bath has to be resolved carefully in order to properly describe the equation
of state of hot QCD above the phase transition. On the other hand, low-energy models of QCD based on the 
NJL model~\cite{Nambu:1961tp} have been extensively used to study the thermodynamics of 
strongly interacting hadronic matter in the vicinity of the phase transition. However, these models are not confining 
and do not include gluodynamics. Therefore phenomenological extensions of these models have been proposed,
see e.~g.~\cite{Meisinger:1995ih,Pisarski:2000eq,Fukushima:2003fw,Braun:2003ii,Ratti:2005jh,Megias:2004hj,Sasaki:2006ww,Schaefer:2007pw}. 
By means of this, the thermodynamics at high temperatures are corrected in the sense that the missing gluon 
contributions are now incorporated. Still, the models are mostly studied in the limit of many colors - 
the large $N_c$ limit. As a consequence the contributions of the scalar fields, i. e. pions in QCD, are not properly taken into 
account. Neglecting the fluctuations of the scalar fields affects universal aspects of the theory in the vicinity of the 
phase transition, such as critical exponents, but also non-universal properties, such as the phase transition 
temperature or the equation of state.

In the present paper, our aim is to study systematically the role of the neglected fluctuations of the scalar fields and 
of the non-trivial momentum dependence of the propagators of the bosonic and fermionic fields. To this end, 
we apply a derivative expansion to a strongly interacting theory of bosons and fermions with $N_c$ colors but only
{\it one} quark flavor. The low-temperature limit of our model is dominated by the presence of a single Goldstone 
boson (pion) rather than three as in QCD with two quark flavors. However, in the spirit of QCD low-energy models
our model is as close as it can be to the large-$N_c$ approximation, where loops involving scalar fields are 
dropped. As we shall discuss below, the large-$N_c$ approximation represents the lowest order of our derivative 
expansion of the effective action.

The paper is organized as follows: In Sect.~\ref{Sec:EffPot} we introduce our model and the non-perturbative 
framework, namely the functional Renormalization Group (RG)~\cite{Wetterich:1992yh}, which we employ 
to compute the thermodynamic properties of our model. The results for the equation of state and the thermal mass of
the scalar fields are then given in Sect.~\ref{sec:results}. In addition we discuss the dependence of the equation of
state on explicit symmetry breaking (i. e. current quark mass) and on the ultraviolet cutoff. 
Our concluding remarks, including a discussion of possible future extensions, are presented in 
Sect.~\ref{sec:conclusions}. 
\section{Functional Renormalization Group and low-energy models}\label{Sec:EffPot}
\subsection{The model and the studied truncations}
Throughout this paper we work in $d=4$ dimensional Euclidean space and employ 
the following ansatz for the effective action for a strongly-interacting theory of bosons and fermions:
\be
\Gamma &=& \int d^4 x\,\left\{ Z_{\psi}\bar{\psi}_{a}\I \partial\!\!\!\slash\,\psi_{a} + \frac{\bar{h}}{\sqrt{2}}\bar{\psi}_{a}(\vec{\tau}\cdot\Phi)\psi_{a}\right.\nn\\
&&\qquad\qquad\qquad\left. + \frac{1}{2}Z_{\phi}\left(\partial_{\mu}\Phi\right)^2 + U(\Phi^2) \right\}\,,\label{Eq:CompleteAction}
\ee
where the index of the fermion fields denotes an internal index, e.~g. a color index in QCD, and runs from $a=1,\dots,N_c$.
The scalar fields are combined in the $O(2)$ vector $\Phi^T=(\sigma,\pi)$ and we have used $\vec{\tau}=(\gamma_5, \I\cdot \mathbf{1})$ 
in order to define a chirally invariant Yukawa interaction. We assume that the bosons are composite degrees of 
fermions and do not carry an internal charge, e.~g. color: $\sigma \sim \bar{\psi}_a\psi_a$ and $\pi \sim  \bar{\psi}_a \gamma_5 \psi_a$.
In the spirit of hadronic models, our ansatz for the effective action represents a low-energy model for QCD 
with one quark flavor. However, we rush to add that our interest is not the study of one-flavor QCD. In this case, we would
have to take the effects of topologically non-trivial gauge configurations into account which are prominent in 
one-flavor QCD since they induce mass-like fermion interactions and break the $U_A(1)$ 
symmetry~\cite{tHooft:1976fv,Shifman:1979uw,Shuryak:1981ff,Schafer:1996wv}. Here, we are interested in an 
investigation of the effect of light bosonic degrees of freedom on the thermodynamics. In this respect, our model with
only one dynamical Goldstone mode allows us to mark out the differences to large $N_c$ studies of presently used 
QCD low-energy models in a simple way. Note that the Goldstone-mode effects become even more pronounced 
with increasing number of quark flavors since the number of Goldstone modes rises quadratically with the number 
of quark flavors. 
\begin{table}[t]
\begin{tabular}{l | l}
\hline\hline
Label &  Approximation \\
 \hline
large $N_c$ & drop bosonic loops \& $\partial_t Z_{\phi}^{\perp} \equiv 0$ \\
LPA  & $\partial_t Z_{\phi}^{\perp} \equiv 0\,, Z_{\phi}^{\parallel} \equiv Z_{\phi}^{\perp}$ \& $\partial_t Z_{\psi}^{\perp} \equiv 0\,, Z_{\psi}^{\parallel} \equiv Z_{\psi}^{\perp}$\\
Trunc. A & $\partial_t Z_{\phi}^{\perp} \neq 0\,, Z_{\phi}^{\parallel} \equiv Z_{\phi}^{\perp}$ \& $\partial_t Z_{\psi}^{\perp} \equiv 0\,, Z_{\psi}^{\parallel} \equiv Z_{\psi}^{\perp}$\\
Trunc. B & $\partial_t Z_{\phi}^{\perp} \neq 0\,, Z_{\phi}^{\parallel} \equiv Z_{\phi}^{\perp}$ \& $\partial_t Z_{\psi}^{\perp} \neq 0\,, Z_{\psi}^{\parallel} \equiv Z_{\psi}^{\perp}$\\
Trunc. C &$\partial_t Z_{\phi}^{\perp} \neq 0\,, \partial_t Z_{\phi}^{\parallel} \neq 0$ \& $\partial_t Z_{\psi}^{\perp} \neq 0\,, \partial_t Z_{\psi}^{\parallel} \neq 0$\\
 \hline\hline
\end{tabular}
\caption{\label{tab:truncation} Specification of the various truncations used to compute the equation of state.}
\end{table}

Let us now discuss the various truncations which we use to study the phase transition temperature and the 
thermodynamics of our model. Our model is built around the standard mean-field ansatz for the effective action 
(large-$N_c$ ansatz), i.~e. 
\be
Z_{\phi}\equiv 0\quad\text{and}\quad Z_{\psi}\equiv 1. \nn
\ee
Such an ansatz has been used extensively for studies of the QCD phase diagram, see e.~g. 
Refs.~\cite{Meyer:2001zp,Buballa:2003qv} and underlies also most of the recent (P)NJL studies of hot 
and dense QCD, see e.~g. Refs.~\cite{Fukushima:2003fw,Megias:2004hj,Ratti:2005jh,Sasaki:2006ww,Schaefer:2007pw}. 

In the following we systematically extend this zeroth-order ansatz in two directions, namely in the number of
derivatives and the number $n$ of external legs of $n$-point functions $\Gamma ^{(n)}$. In order to 
study spontaneous symmetry breaking which is indicated by a non-trivial minimum of the order-parameter 
potential $U(\Phi^2)$ in Eq.~\eqref{Eq:CompleteAction}, we expand the potential in powers of $\Phi^2$. This results in RG
flow equations for the mesonic $n$-point functions. On the other hand, we perform a derivative expansion which
renders the $n$-point functions momentum dependent. Note that the latter is indispensable for a connection of the
high- and low-momentum regime of QCD~\cite{Gies:2002hq,Pawlowski:2005xe,Braun:2008pi}, 
see also Refs.~\cite{Gies:2001nw,Floerchinger:2009uf}.

Next to the zeroth-order approximation (large $N_c$) one needs to include kinetic terms for the meson fields. The
minimal truncation which allows for an inclusion of meson loops is given by the so-called Local Potential 
Approximation (LPA), i.~e.
\be
Z_{\phi} \equiv 1 \quad\text{and}\quad Z_{\psi} \equiv 1\,. \nn
\ee
This truncation has been used, e.~g., in Refs.~\cite{Schaefer:1999em,Braun:2003ii,Schaefer:2004en,Braun:2005fj} 
for a study of the quark-meson model at finite temperature and density. It turns out that the LPA represents indeed 
already a major improvement with respect to the quality of the critical exponents. At this point, we would like to 
remark that the quality of critical exponents can be considered as a measure of how well the dynamics at the phase 
transition are captured. 

Next to LPA, we consider a truncation in which we allow for a running of the scalar wave-function renormalization,
but keep the wave-function renormalization of the fermions fixed, i. e. 
\be
Z_{\phi}(k\!=\!\Lambda)\ll 1\,,&& \partial_t Z_{\phi}\neq 0\,,\nn\\
Z_{\psi}(k\!=\!\Lambda)= 1\,,&& \partial _t Z_{\psi} \equiv 0\,,\nn
\ee
where $t=\ln(k/\Lambda)$ and $k$ denotes the IR (infrared) cutoff scale of our functional RG approach and $\Lambda$ 
denotes the UV (ultraviolet) cutoff, see Subsec.~\ref{subsec:rgfinitet} for details. 
By means of this approximation, vertices involving at least one boson field become momentum dependent. 
Thereby the quality of the results is already improved considerably which can be read off from the quality of the critical 
exponents\footnote{The derivative expansion can be continued systematically by e.~g. including terms of the 
form $Y_k(\Phi\partial_{\mu} \Phi)^2$ in our truncation~\cite{VonGersdorff:2000kp}.}, see~e.~g. 
Refs.~\cite{Tetradis:1993ts,Berges:1997eu,Berges:2000ew}. We shall denote this as truncation A from now on.

In our truncation B, we go beyond this approximation and allow for a running of the wave-function renormalizations of 
the fermion fields as well, i.~e. 
\be
Z_{\phi}(k\!=\!\Lambda) \ll 1\,,&& \partial_t Z_{\phi}\neq 0\,,\nn\\
Z_{\psi}(k\!=\!\Lambda)= 1\,,&& \partial _t Z_{\psi}\neq 0\,.\nn
\ee
Such a truncation renders all vertices momentum dependent. 

At finite temperature the wave-function renormalizations longitudinal ($\parallel$) 
and transversal ($\perp$) to the heat-bath obey in general a different RG running.
In order to study this difference, we have to replace the part of the truncation~\eqref{Eq:CompleteAction} 
involving derivatives according to
\be
&&Z_{\psi}\bar{\psi}_{a}\I \partial\!\!\!\slash\,\psi_{a} + \frac{1}{2}Z_{\phi}\left(\partial_{\mu}\Phi\right)^2 \nn\\
&&\;\;\longrightarrow
Z_{\psi}^{\parallel}\bar{\psi}_{a}\I \gamma_0 \partial_0\,\psi_{a} + \frac{1}{2}Z_{\phi}^{\parallel}\left(\partial_{0}\Phi\right)^2\nn\\
&&\qquad\qquad +Z_{\psi}^{\perp}\bar{\psi}_{a}\I  \gamma_i \partial_i\,\psi_{a}
+ \frac{1}{2}Z_{\phi}^{\perp}\left(\partial_{i}\Phi\right)^2\,,
\ee
where we identify $Z_{\psi,\phi}^{\perp}$ with $Z_{\psi,\phi}$ in both truncation A and B but allow for a different
running of  $Z_{\psi,\phi}^{\perp}$ and $Z_{\psi,\phi}^{\parallel}$. We shall denote this as truncation C.
The various truncations alongside with their labels are briefly summarized in Tab.~\ref{tab:truncation}.

As mentioned above, we expand the scalar potential $U(\Phi^2)$ (order-parameter potential) 
in a power series of $\Phi^2$ up to order $\Phi^4$ and drop all higher terms~\cite{Braun:2007td,Braun:2008sg,Nakano:2009ps}: 
\be
U(\Phi^2)\equiv U(\rho,\sigma)&=&\sum_{\alpha=0}^{\infty}\, \frac{\bar{\lambda}_{\alpha}}{\alpha !} \left(\rho - \rho_0\right)^{\alpha} - c\sigma\nn\\
&\approx & \bar{\lambda}_0 + \bar{\lambda}_1 \left(\rho \!-\! \rho_0\right) + \frac{\bar{\lambda}_2}{2} \left(\rho \! -\! \rho_0\right)^2 \!-\! c\sigma\,,\nn\\
\label{eq:potansatz}
\ee
where $\rho=(1/2)\Phi^2$, $\Phi _0^{T} =(\sigma_0,0)$ and the parameter $c$ is a measure for the strength of
explicit symmetry breaking. Note that such an expansion represents a systematic expansion in $n$-point functions 
$\Gamma ^{(n)}$ where $n=2\alpha$ defines the number of external legs and is well under control. 
In fact, the quality of an expansion of $U(\Phi^2)$ in powers of $\Phi^2$ has been studied quantitatively in 
Ref.~\cite{Tetradis:1993ts} at vanishing temperature and for the proper-time RG in LPA at finite temperature in 
Ref.~\cite{Papp:1999he}. In Ref.~\cite{Schaefer:2004en}, the order-parameter potential $U$ has been computed 
in LPA for a quark-meson model with two quark flavors within a RG approach 
without making use of a Taylor expansion in~$\Phi$. Note that we also solve the RG flow in the 
large-$N_c$ approximation by making use of the expansion~\eqref{eq:potansatz} of the order-parameter potential $U$, even though it is 
straightforward to solve the flow in this case without truncating the power series in Eq.~\eqref{eq:potansatz}, 
see e.~g. Ref.~\cite{Meyer:2001zp}. In our case a Taylor expansion of the potential is justified since (i) we want to 
focus on the role of the momentum dependence of the vertices and (ii) have to ensure comparability of our results from 
our various truncations.

From the condition $\partial U/\partial \sigma =0$ evaluated at $\sigma=\sigma_0$, we find that the RG flow of the 
coupling $\bar{\lambda}_1$ and the minimum $\sigma_0(k)$ are related:
\be
\bar{\lambda}_1(k) \sigma_0(k) = c\,.\label{eq:min_cond2}
\ee 
This condition keeps the minimum at $(\sigma, \pi) = (\sigma_0(k), 0)$. The flow equation of the minimum $
\sigma_0(k)$ is thus related to the flow of the coupling $\bar{\lambda}_1$ in a simple way.
The initial conditions for the various couplings in Eq.~\eqref{Eq:CompleteAction} at the ultraviolet scale $\Lambda$ 
are discussed in Sec.~\ref{sec:results}. 
\smallskip

\subsection{RG flow at finite temperature}\label{subsec:rgfinitet}
For our derivation of the RG flow equations of the couplings, we employ the Wetterich 
equation~\cite{Wetterich:1992yh}:
\be
\partial _t \Gamma_{k} [\chi] 
= \frac{1}{2} \STr\left\{\left[\Gamma _{k} ^{(1,1)}[\chi] + R_k\right]^{-1}
\!\cdot\!\left(\partial _t R_k\right)\right\}
\ee
with
\be
\Gamma _{k} ^{(1,1)}[\chi]=\fderL{\chi ^{T}}\Gamma _{k} [\chi]\fderR{\chi}\,,\label{eq:Flow_EffAction}
\ee
where $t=\ln k/\Lambda$ and $\Lambda$ is the UV cutoff scale. Here, $\chi$ represents a vector in field space 
and is defined as
\be
\chi ^T \equiv \chi ^T (-q):=\left(\psi ^T (-q),\bar{\psi} (q),\Phi_1 (-q),\Phi_2 (-q)\right)\nn
\ee
and
\be
\chi\equiv\chi(q):= \left(\begin{array}{c}
 {\psi (q)}\\
{\bar{\psi}^T(-q)}\\
{\Phi_1 (q)}\\
{\Phi_2 (q)}
\end{array}
\right)\,.\nn
\ee
Thus, $\Gamma _{k} ^{(1,1)}[\chi]$ is matrix-valued in field space and so is the regulator function $R_k$.
In this work, we employ a 3$d$ optimized regulator function which is technically advantageous for studies at
finite temperature~\cite{Braun:2003ii,Litim:2006ag,Blaizot:2006rj}. The quality of such a 3$d$ regulator in the 
limit of vanishing temperature and chemical potential has been estimated by computing critical exponents 
of $O(N)$ models~\cite{Litim:2001hk} and comparing them to those optained with an optimized regulator 
in 4$d$ space-time~\cite{Litim:2000ci,Litim:2001up}. Details on the regularization can be found in 
App.~\ref{sec:thresholdfcts}. In the following we give the RG flow 
equations in a way which does not dependent on the details of our 3$d$ regularization. Reviews on and 
introductions to the functional RG and its application to gauge theories can be found in 
Refs.~\cite{Litim:1998nf,Berges:2000ew,Polonyi:2001se,Pawlowski:2005xe,Gies:2006wv}.

Inserting our ansatz for the effective action~\eqref{Eq:CompleteAction} into the flow equation~\eqref{eq:Flow_EffAction}, 
we find the flow equation for the order-parameter potential $U$:
\be
\partial _t U &=& \frac{8v_3}{3}k^4 \Bigg\{  
\frac{\hat{z}_{\phi}^{-1/2}}{\sqrt{1\!+\! m_{\sigma}^2}}\left(\frac{1}{2}\!+\! n_B  (\hat{z}_{\phi},m_{\sigma})\right) \nn\\
&&\qquad\quad + \frac{\hat{z}_{\phi}^{-1/2}}{\sqrt{1\!+\! m_{\pi}^2}} \left(\frac{1}{2}\!+\! n_B  (\hat{z}_{\phi},m_{\pi})\right)\nn\\
&&\qquad\quad\; -\, \frac{2 N_c \hat{z}_{\psi}^{-1}}{\sqrt{1\!+\! m_{q}^2}} \left(\frac{1}{2}\!-\! n_F (\hat{z}_{\psi},m_{\psi})\right)
\!\!\Bigg\},
\label{eq:pot_flow}
\ee
where $v_3=1/(8\pi^2)$, $\hat{z}_{\phi}=Z_{\phi}^{\parallel}/Z_{\phi}^{\perp}$, $\hat{z}_{\psi}=Z_{\psi}^{\parallel}/Z_{\psi}^{\perp}$ 
and
\be
n_B(\hat{z},m)&=&\left(\E ^{\frac{\sqrt{k^2 + m^2}}{\hat{z}^{1/2}T}}-1\right)^{-1}\,,\\
n_F(\hat{z},m)&=&\left(\E ^{\frac{\sqrt{k^2 + m^2}}{\hat{z}T}}+1\right)^{-1}\label{eq:dist_functions}
\ee
denote the Bose-Einstein and Fermi-Dirac distribution function, respectively. The renormalized
masses of the bosons and fermions are given by
\be
m_{\sigma}^2  &=&\frac{1}{Z_{\phi}^{\perp} k^2}\left( \frac{\partial U}{\partial \rho} + 2 \rho  \frac{\partial ^2 U}{\partial \rho^2}\right)\,,\\
m_{\pi}^2  &=&\frac{1}{Z_{\phi}^{\perp} k^2}\left( \frac{\partial U}{\partial \rho}\right)\,,\\
m_{\psi}^2 &=& \frac{\bar{h}^2 \rho}{(Z_{\psi}^{\perp})^2 k^2} = h^2 \left(\frac{Z_{\phi}^{\perp}\rho}{ k^2}\right)\,.
\ee
Here we have introduced the renormalized Yukawa coupling $h=\bar{h}/(Z^{\perp}_{\psi}(Z_{\phi}^{\perp})^{1/2})$. 
Note that the large-$N_c$ limit can be deduced from Eq.~\eqref{eq:pot_flow} by dropping the bosonic loops as 
well as the running of the wave-function renormalizations.

In Eq.~\eqref{eq:pot_flow} and~\eqref{eq:dist_functions} we observe that the ratio of the wave-function renormalizations 
$\hat{z}_{\phi}=Z_{\phi}^{\parallel}/Z_{\phi}^{\perp}$ and $\hat{z}_{\psi}=Z_{\psi}^{\parallel}/Z_{\psi}^{\perp}$ 
effectively rescale the (dimensionless) temperature $\tilde{t}=T/k$ in the loop diagrams.

The anomalous dimensions,
\be
\eta_{\phi}^{\parallel,\perp}=-\partial_t \ln Z_{\phi}^{\parallel,\perp}\,,
\ee
associated with the scalar wave-function renormalizations longitudinal $(\parallel)$ and transversal $(\perp)$ to 
the heat-bath are given by
\be
\eta_{\phi}^{\perp} &=& \frac{4 v_3}{3}N_c \Big[
h^2 {\mathcal M}_{4,\perp}^{(F)}(\tilde{t},m_{\psi}^2) 
+ 4 \kappa h^4  {\mathcal M}_{2,\perp}^{(F)}(\tilde{t},m_{\psi}^2) \nn\\
&& \qquad\qquad + 4\frac{\kappa\lambda_{\phi}^2}{N_c} {\mathcal M}_{2,2,\perp}^{(B)}(\tilde{t},m_{\sigma}^2,m_{\pi}^2)\Big]\,,\\
\eta_{\phi}^{\parallel} &=& 
 \frac{4 v_3}{\hat{z}_{\phi}}N_c \Big[
h^2 {\mathcal M}_{4,\parallel}^{(F)}(\tilde{t},m_{\psi}^2) 
+ 4 \kappa h^4  {\mathcal M}_{2,\parallel}^{(F)}(\tilde{t},m_{\psi}^2) \nn\\
&& \qquad\qquad + 2\frac{\kappa\lambda_{\phi}^2}{N_c} {\mathcal M}_{2,2,\parallel}^{(B)}(\tilde{t},m_{\sigma}^2,m_{\pi}^2)\Big]\,,
\ee
where $\kappa = Z_{\phi}^{\perp} \rho_0/k^2$ and $\lambda_{\phi}=\bar{\lambda}_2/(Z_{\phi}^{\perp})^2$. 
The anomalous dimensions,
\be
\eta_{\psi}^{\parallel,\perp}=-\partial_t \ln Z_{\psi}^{\parallel,\perp}\,,
\ee
of the fermion fields read
\be
\eta_{\psi}^{\perp} &=& \frac{4v_3}{3} h^2 \Big[ {\mathcal M}_{1,2,\perp}^{(FB)} (\tilde{t},m_{\psi}^2,m_{\pi}^2) \nn\\
&&\qquad\qquad + {\mathcal M}_{1,2,\perp}^{(FB)} (\tilde{t},m_{\psi}^2,m_{\sigma}^2)  \Big]\,,\\
\eta_{\psi}^{\parallel} &=& \frac{4v_3}{\hat{z}_{\psi}} h^2 \Big[ {\mathcal M}_{1,2,\parallel}^{(FB)} (\tilde{t},m_{\psi}^2,m_{\pi}^2) \nn\\
&&\qquad\qquad + {\mathcal M}_{1,2,\parallel}^{(FB)} (\tilde{t},m_{\psi}^2,m_{\sigma}^2)  \Big]\,.
\ee
The threshold functions ${\mathcal M}$ are defined in Ref.~\cite{Braun:2008pi} and App.~\ref{sec:thresholdfcts} and 
describe the decoupling of massive and thermal modes. The RG flows of the ratios 
$\hat{z}_{\phi}=Z_{\phi}^{\parallel}/Z_{\phi}^{\perp}$ and $\hat{z}_{\psi}=Z_{\psi}^{\parallel}/Z_{\psi}^{\perp}$ read
\be
\hat{\eta}_{\phi}=-\frac{\partial_t \hat{z}_{\phi}}{\hat{z}_{\phi}}=\eta_{\phi}^{\parallel}-\eta_{\phi}^{\perp}\,,\label{eq:zflow}\\
\hat{\eta}_{\psi}=-\frac{\partial_t \hat{z}_{\psi}}{\hat{z}_{\psi}}=\eta_{\psi}^{\parallel}-\eta_{\psi}^{\perp}\,.\label{eq:zflow}
\ee
Since $\eta_{\phi}^{\parallel} =\eta_{\phi}^{\perp}$ and $\eta_{\psi}^{\parallel} =\eta_{\psi}^{\perp}$ for 
$T/k=\tilde{t}=0$, which is a property of the threshold functions ${\mathcal M}$, we have $\hat{\eta}_{\phi}=0$ for 
$\tilde{t}=0$ as expected for a Poincare-invariant theory.

The RG flow of the Yukawa coupling can be calculated along the lines of Ref.~\cite{Braun:2008pi}. We find 
\be
\partial_t h^2 &=& (\eta_{\phi}^{\perp} +2\eta_{\psi}^{\perp}) h^2
- 4 v_3 h^4 \Big[
L_{1,1} ^{(FB)}(m_{\psi}^2,0,m_{\pi}^2)\nn\\
&& \qquad\quad - L_{1,1} ^{(FB)}(\tilde{t},m_{\psi}^2,0,m_{\sigma}^2)\Big]
\,.\label{eq:hflow}
\ee
The threshold function $L_{1,1} ^{(FB),(4)}$ can be found in Ref.~\cite{Braun:2008pi}.
In the limit $\tilde{t}=T/k\to 0$, we recover the perturbative result $\partial _t h^2 = ((N_c+1)/(8\pi ^2))h^4$, 
see e.~g.~Ref.~\cite{Gies:2002hq}. In the chirally symmetric regime ($m_{\sigma}$=$m_{\pi}$), we find that 
the RG flow of the Yukawa coupling is governed by the IR attractive fixed point $h^{*}=0$ for $\tilde{t}=T/k\to 0$. 
For $\tilde{t}\to\infty$, we have $\beta_{h^2}\equiv \partial_t h^2\to 0$ which is due to the decoupling of the fermions 
for $k\lesssim T$. Note that it is special to our model with just one fermion species ($N_f=1$) that
the running of the Yukawa coupling is only driven by the anomalous dimensions in the symmetric regime. 
For example, in QCD with more than one quark flavor we would find additional contributions to the RG flow of 
the Yukawa coupling since the contributions from the Goldstone modes in the symmetric regime are not canceled 
completely by the ones of the radial mode~\cite{Jungnickel:1995fp,Gies:2009sv}. In the regime with broken $O(2)$ 
symmetry in the ground state, these additional terms emerge when $m_{\sigma}\neq m_{\pi}$.

Let us close this section with some general remarks on the RG flow of  $Z_{\psi,\phi}^{\parallel}$ and 
$Z_{\psi,\phi}^{\perp}$ at vanishing temperature. As mentioned above, we employ a 3$d$ regulator 
function in order to derive the flow equations. Thus, our regularization breaks necessarily the $O(d)$ symmetry 
in the derivative terms of our truncation, i. e.~$\partial_t Z_{\psi,\phi}^{\parallel} \neq \partial_t Z_{\psi,\phi}^{\perp}$
even for $T/k \to 0$. In LPA and in large-$N_c$ approximations, this problem does not appear since the non-trivial 
momentum dependence of the propagators is 
neglected~\cite{Braun:2003ii,Schaefer:2004en,Litim:2006ag,Blaizot:2006rj}. In studies beyond LPA, we have to 
deal with the broken Poincare-invariance at vanishing temperature due to the choice of our $3d$ regulator 
function\footnote{This issue does not occur if one applies a $4d$ regulator function.}. 
In principle one can solve this problem by taking care of the symmetry violating terms with the aid of 
the corresponding Ward identities. Equivalently, we can choose the initial conditions for the RG flow equations 
such that one finds $Z_{\psi,\phi}^{\perp}=Z_{\psi,\phi}^{\parallel}$ for $k\to 0$ and $T\to 0$. 
In the present paper, we adjust the initial conditions properly in order to deal with this 
circumstance\footnote{From a field-theoretical point of view the adjustment of the initial conditions
of a given truncation means nothing else than adding appropriate counter-terms such that the theory remains 
Poincare-invariant for $k\to0$ and $T\to0$.}. At finite temperature, the breaking of the $O(d)$ symmetry in 
momentum space due to our choice of the regulator function does not harm our study since this symmetry is 
broken anyway. However, the choice of our regulator function allows us to perform the Matsubara sums analytically,
which simplifies the numerics considerably and justifies our choice for the regulator function.
\section{Equation of state and thermal mass}\label{sec:results}
Let us first discuss our choice for the initial conditions. Since we are interested in studying how the 
momentum dependence of the propagators and vertices affects the phase-transition temperature
and the thermodynamic pressure,  we adjust the parameters (initial conditions at the UV scale $\Lambda$) 
of the model such that we obtain the same values for the IR observables for all truncations in this work. 
To be specific, we choose $\Lambda=2\,\text{GeV}$ and 
$f_{\pi}\equiv\sqrt{\rho_0}=0.030\,\text{GeV}$, $m_q=0.088\,\text{GeV}$ and $Z_{\psi,\phi}^{\perp}=Z_{\psi,\phi}^{\parallel}$
at $k=0$ and $T=0$. Moreover we use $N_c=3$ in the numerical evaluation of the RG flow equations.
Our choice for the parameters ensures that (i) the UV cutoff $\Lambda$ is 
much larger than the physically relevant scales in order to avoid cutoff effects in the high-temperature
regime\footnote{While the UV cutoff $\Lambda$ can be chosen almost arbitrarily in LPA and in our large-$N_c$ 
approximation, the possible choices for the initial conditions (UV parameters) for a given set of IR observables in 
truncations beyond LPA is restricted by the running of the Yukawa coupling. In this sense, these models have some
predictive power even at vanishing temperature~\cite{Jungnickel:1995fp}.} ($T>T_{\rm{c}}$) and (ii) that
dimensionless ratios of observables are close to the ones in QCD with $N_f=2$ and $N_f=2+1$ quark flavors. 
Indeed, our choice for the parameters yields $m_q/f_{\pi}\approx 3$. Since the actual dimensionful values for the
critical temperature is of no interest in this paper, we give all results in units of $f_{\pi}$. This enables us to
compare the results from the various truncations in a simple manner.

\begin{table}[t]
\begin{tabular}{l | c | c | c | c}
\hline\hline
Truncation &  $T_{\rm{c}}/f_{\pi}$  &  $T_{\text{max}}/f_{\pi}$ & $T_{\text{max}}/T_{\rm{c}}$ & $k_{\chi\text{SB}}^{T=0}/f_{\pi}$ \\
 \hline
large $N_c$ & 2.87 & --- & --- & 8.96\\
LPA  & 2.12 & 2.35 & 1.11 & 8.93 \\
Trunc. A & 1.96 & 2.12 & 1.08 & 8.24\\
Trunc. B & 2.06 &2.26 & 1.10 & 8.85\\
Trunc. C & 1.93& 2.12 & 1.09 & 8.78\\
 \hline\hline
\end{tabular}
\caption{\label{tab:crit_temp} Results for the phase transition temperature $T_{\rm{c}}$, the position of the 
maximum in the pressure, $T_{\text{max}}$, and the chiral symmetry breaking scale $k_{\chi\text{SB}}$ at 
vanishing temperature. All values are given in units of the value of the order parameter 
$f_{\pi}=\sqrt{\rho_0}=\sigma_0/\sqrt{2}=30\,\text{MeV}$ at vanishing temperature. 
}\label{tab:results}
\end{table}

The results for the phase-transition temperature $T_{\rm{c}}$ for the various truncations are listed in 
Tab.~\ref{tab:results}. We observe that the phase transition temperature decreases when we take bosonic 
loops into account and resolve the momentum dependence of the vertices. Since the Goldstone boson plays a
prominent role at the phase transition, it is intuitively clear that the inclusion of these effects affect the dynamics 
near the phase transition. Moreover, Goldstone bosons tend to restore the symmetry while fermions tend to build 
up a condensate and thereby break the symmetry of the ground state. However, the anti-periodic boundary conditions for 
the fermions in Euclidean time direction lead to a suppression of the fermionic modes in the vicinity and above the 
phase transition due to the absence of a zero mode.

\begin{figure}[t]
\begin{center}
\includegraphics[scale=0.68,clip=]{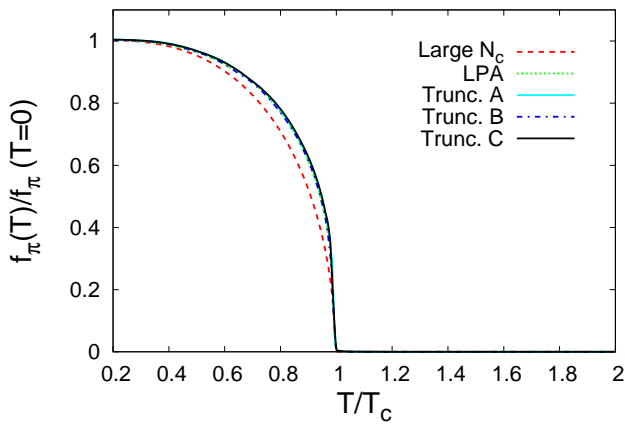}
\end{center}
\vspace*{-4mm}
\caption{(Color online)
Order parameter divided by its zero-temperature value as a function of $T/T_{\rm{c}}$ as obtained from the various
different truncations for the momentum dependence. For all truncations beyond large-$N_c$, the results for the 
order parameter are indistinguishable on the shown resolution. The different shape of the large-$N_c$ result is 
due to the fact that the value of the corresponding critical exponent differs from the exact value.
}\label{fig:vev}
\vspace*{-4mm}
\end{figure}

In Fig.~\ref{fig:vev} the influence of the bosonic fluctuations is visible in a different observable, namely the 
behavior of the order parameter near the phase transition which can be measured in terms of a 
universal critical exponent~$\beta$, $f_{\pi}\sim|T-T_{\rm{c}}|^{\beta}$. We observe that the slope of the order
parameter at the phase transition found in our large-$N_c$ study deviates clearly from the slope found in 
truncations including bosonic fluctuations. For example, we find 
$\beta_{\text{large }N_c}= 0.500\pm 0.001$ and 
$\beta_{\text{LPA}}= 0.349\pm 0.011$. Our large-$N_c$ value agrees nicely with the
analytically known value,  $\beta_{\text{large }N_c}^{\text{exact}}=1/2$. The LPA value is in agreement  
with the corresponding values found in Refs.~\cite{Litim:2002cf} and~\cite{Tetradis:1993ts}. 
The error bars arise from a fit to our numerical data for the order parameter.

\begin{figure}[t]
\begin{center}
\includegraphics[scale=0.68,clip=]{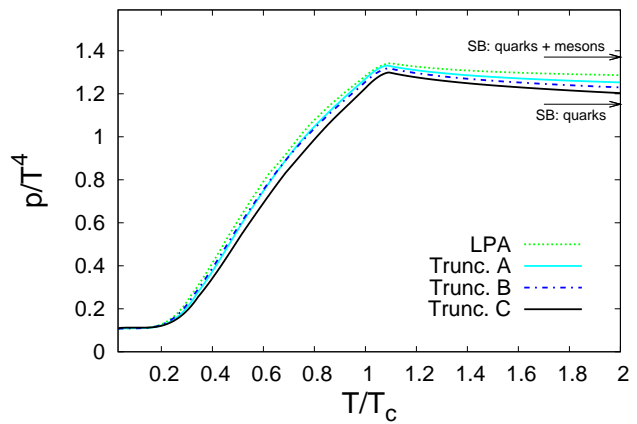}
\end{center}
\vspace*{-4mm}
\caption{(Color online)
Pressure as a function of $T/T_{\rm{c}}$ as obtained from the different truncations of the momentum dependence of the 
propagators and vertices. We observe a maximum in the pressure function located slightly above $T_{\rm{c}}$. 
For high temperatures, we observe that the Stefan-Boltzmann (SB) limit of a free gas of fermions (quarks) 
is approached from above. 
}\label{fig:pressure}
\vspace*{-4mm}
\end{figure}

Next we discuss our results for the pressure. The flow equation for the pressure is obtained from the negated 
right-hand side of the flow equation~\eqref{eq:pot_flow} for the order-parameter potential evaluated at the 
physical ground state $\Phi_0$. In Fig.~\ref{fig:pressure}, we show our results for the pressure scaled by 
$T^4$ as a function of $T/T_{\rm{c}}$ as obtained from LPA as well as from the truncations A-C; we shall 
discuss the large-$N_c$ limit below. Since we are considering the limit of vanishing external linear symmetry 
breaking (chiral limit, $c\to 0$ in Eq.~\eqref{eq:potansatz}), we observe that the pressure approaches the 
Stefan-Boltzmann (SB) limit of one massless boson for small temperatures, $p/T^4 \to \pi^2/90$ for $T\to 0$.
When we increase the temperature starting from $T=0$, we find that the pressure increases since the mass of 
the fermions, $m_q=h f_{\pi}$, decreases, see also Fig.~\ref{fig:vev}. At $T=T_{\rm{c}}$ the fermions become 
massless while the bosons acquire a thermal mass $m_{\phi}=m_{\sigma}=m_{\pi}$. However, the pressure neither takes its 
maximal value at $T_{\rm{c}}$ nor does it agree with the Stefan-Boltzmann limit of a free gas of massless fermions. 
On the contrary, we observe that the pressure function reaches a maximum at $T_{\text{max}}>T_{\rm{c}}$ 
and approaches the Stefan-Boltzmann limit of a gas of free massless quarks for $T>T_{\text{max}}$ slowly from above.
We would like to stress that this is a substantial difference to the findings in a large-$N_c$ approximations as we 
shall see below. At $T=T_{\text{max}}$, we find 
\be
\frac{p}{T^4}\Big|_{T=T_{\text{max}}}\approx 4 N_c \frac{7}{8}\frac{\pi^2}{90} + 2  \frac{\pi^2}{90}\,,
\ee
which is the pressure of a gas of $4 N_c$ massless fermions  and $2$ massless bosons. This result 
can be easily generalized to the case with $N_f$ fermion species. We expect that the maximal pressure of a QCD 
low-energy model with $N_f \geq 2$ quark flavors is given by\footnote{Our observation of the existence of a 
maximum in the pressure function is in accordance with an estimate for the pressure function in Ref.~\cite{Schaefer:1999em}
obtained from studying a gas of non-interacting quarks and mesons with temperature-dependent masses.}
\be
&&\frac{p}{T^4}\Big|_{T_{\text{max}},N_f\geq 2}\approx 4 N_c N_f \frac{7}{8}\frac{\pi^2}{90} + N_f^2 \frac{\pi^2}{90}\nn\\
&&\quad\Rightarrow 
\frac{p}{p_{\text{free\ quarks}}}\Big|_{T=T_{\text{max}},N_f\geq 2}\approx 1+ \frac{2}{7}\frac{N_f}{N_c}\,.\label{eq:max2flavor}
\ee
In the second line we have divided our estimate for the ratio of the pressure  at $T=T_{\text{max}}$ with the 
Stefan-Boltzmann limit of a gas of free massless quarks.

\begin{figure}[t]
\begin{center}
\includegraphics[scale=0.650,clip=]{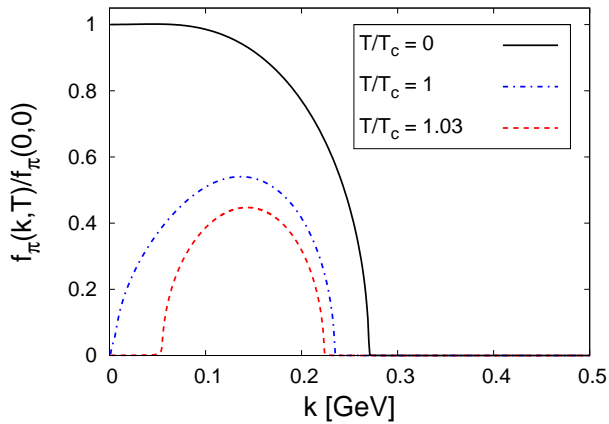}
\end{center}
\vspace*{-4mm}
\caption{(Color online)
RG flow of the pion decay constant in LPA for $T/T_{\rm{c}}= 0,\,1,\,1.03$. We observe that a regime exists in 
which the pion decay constant is finite even though the system is the phase with restored chiral symmetry for $k\to 0$. 
Note that the behavior of $f_{\pi}$ is qualitatively the same for the truncations A-C.
}\label{fig:vev_flow}
\vspace*{-4mm}
\end{figure}

The observation that $T_{\text{max}}>T_{\rm{c}}$ reflects the fact that chiral symmetry is not restored on all scales of 
the RG flow for $T_{\rm{c}} \leq T \leq T_{\text{max}}$ but broken on intermediate scales $k>0$. This is depicted in
Fig.~\ref{fig:vev_flow} where we show the pion decay constant $f_{\pi}$, i. e. the order parameter, as a function of
the RG scale $k$ in the temperature regime $T_{\rm{c}} \leq T \leq T_{\text{max}}$. The behavior of $f_{\pi}$ 
displays the interplay of the bosons and fermions in the RG flow: The fermions dominate the dynamics in the RG 
flow and tend to break chiral symmetry for $k/T\gg 1$. Below the chiral symmetry breaking scale $k_{\chi \text{SB}}$
the fermions acquire then a mass and therefore decouple from the flow. Moreover all fermionic modes acquire a thermal 
mass $\sim T$. In contradistinction to the fermions, the bosonic fields have a thermal zero mode. Thus the RG flow 
is governed by the Goldstone boson for $k/T \lesssim 1$ which drives the system towards the regime with restored 
chiral symmetry. Since the IR regulator scale $k$ introduces an intrinsic length scale $\sim 1/k$, we conclude that we 
have "local order" for  $T_{\rm{c}} \leq T \leq T_{\text{max}}$, i. e. the system is ordered on spatial 
domains\footnote{We would like to remark that the inverse IR regulator scale $1/k$ can be roughly associated 
with the side length $L$ of the spatial simulation volume in lattice QCD simulations, see also Ref.~\cite{Braun:2004yk}.} 
of length scales $\sim 1/k$. This implies that the bosons contribute still significantly to the pressure for these temperatures. 
On the other hand, the fermions give a significant contribution as well in this regime since their mass is zero for $k\to 0$. For 
$T>T_{\text{max}}$, the system remains in the symmetric regime on all RG scales. Thus we have "global disorder" 
for these temperatures. Let us further discuss this phenomenon: In a Landau-Ginzburg type description of phase 
transitions, symmetry breaking is indicated by a negative squared mass parameter (i. e. by a finite vacuum 
expectation value of the order-parameter potential). In an RG approach this parameter is scale-dependent. Usually 
we are only interested in the shape of the potential in the limit $k\to 0$. For $T>T_{\text{max}}$ the squared mass 
parameter of the Landau-Ginzburg potential remains positive on all scales. On the other hand, the squared mass 
parameter becomes negative on intermediate scales for $T_{\rm{c}} \leq T \leq T_{\text{max}}$ but becomes positive
again in the physical limit $k\to 0$, see also Fig.~\ref{fig:vev_flow} where the regimes of positive mass squared 
correspond to a vanishing order parameter. Below the phase transition temperature ($T<T_{\rm{c}}$) the squared 
mass parameter becomes negative at the chiral symmetry breaking scale $k_{\chi\text{SB}}$ and stays negative 
even for $k\to 0$. 

Note that $T_{\rm{c}}$ and $T_{\text{max}}$ differ only by roughly $10\%$, see Tab.~\ref{tab:crit_temp}. Therefore 
the phase transition temperature can indeed be estimated reasonably well by just seeking for the temperature for 
which the squared mass parameter of a Landau-Ginzburg--type potential stays positive on all scales. This strategy
has been applied in Refs.~\cite{Braun:2005uj,Braun:2006jd} in order to estimate the chiral phase transition in 
QCD from RG flows involving gluonic degrees of freedom.
\begin{figure}[t]
\begin{center}
\includegraphics[scale=0.68,clip=]{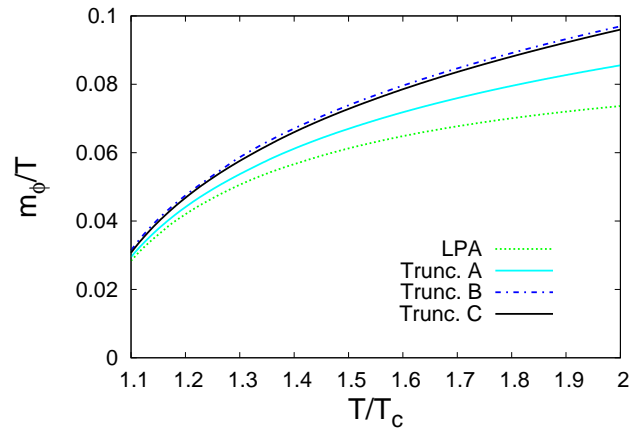}
\end{center}
\vspace*{-4mm}
\caption{(Color online)
Thermal mass $m_{\phi}$ of the scalar fields divided by the temperature as a function of $T/T_{\rm{c}}$ as obtained from the 
different truncations for the momentum dependence. The increase of the thermal mass goes alongside with the decrease
of the pressure function for the shown temperatures.
}\label{fig:thermalmass}
\vspace*{-4mm}
\end{figure}
In Fig.~\ref{fig:pressure} and Tab.~\ref{tab:results}, we also observe that the position of the maximum of the pressure
is almost independent of the truncation. For $T>T_{\text{max}}$ we find that the Stefan-Boltzmann limit of a free gas
of massless quarks is approached faster for truncations in which the momentum dependence of the vertices has
been resolved more accurately. In general, the behavior of the pressure function for $T>T_{\rm{c}}$ reflects the 
behavior of the thermal bosonic mass $m_{\phi}$. In Fig.~\ref{fig:thermalmass} we show $m_{\phi}/T$ 
as a function of $T/T_{\rm{c}}$ as obtained from LPA as well as from the truncations A-C. For asymptotically high
temperatures we expect that $m_{\phi}/T =\text{const.}$ according to perturbation theory~\cite{Dolan:1973qd}. We
find that this limit is not yet reached for the temperature range studied here. Moreover, we observe a difference 
between the results from LPA and from truncation~C. The difference between the results from truncation~B and 
truncation~C, where we have also resolved the difference in the running of the wave-function renormalizations in 
the Euclidean time and space directions, is small for $T/T_{\rm{c}}\lesssim 2$. This represents 
an {\it a posteriori} justification for many studies where $Z_{\phi,\psi}^{\parallel}\equiv Z_{\phi,\psi}^{\perp}$ has 
been assumed, see e.~g. Refs.~\cite{Berges:1997eu,Berges:2000ew,Bohr:2000gp,Braun:2008pi}. 
For a high-accuracy study of critical dynamics, however, our results suggest that higher-order derivatives may 
still play a role in the vicinity of the phase transition temperature.
Therefore an inclusion of the full momentum dependence of the vertices is important for such a high-accuracy 
study. At vanishing temperature, the momentum dependence has been fully resolved in a study of critical 
exponents of $O(N)$ models which has led to accurate predictions~\cite{Benitez:2009xg}.

\begin{figure}[t]
\begin{center}
\includegraphics[scale=0.68,clip=]{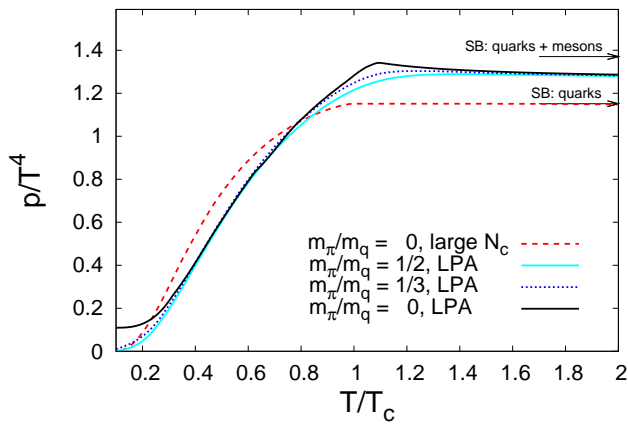}
\end{center}
\vspace*{-4mm}
\caption{(Color online)
Pressure as a function of $T/T_{\rm{c}}$ for a large-$N_c$ approximation of our RG flow equations
and for LPA for various ratios $m_{\pi}/m_q=0,\frac{1}{3},\frac{1}{2}$. 
}
\label{fig:largeN}
\vspace*{-4mm}
\end{figure}

Let us now discuss the results obtained from our large-$N_c$ approximation. In Fig.~\ref{fig:largeN} we compare
our results for the pressure from a large-$N_c$ approximation in the chiral limit with the results 
from LPA for $m_{\pi}/m_q=0,\,1/3,\,1/2$. In particular $m_{\pi}/m_q=1/3$ resembles the situation in 
QCD with two quark flavors, where we have pions with a mass of roughly $130\,\text{MeV}$ and
constituent quarks with a mass of roughly $330\,\text{MeV}$. In large $N_c$ and LPA with $m_{\pi}/m_q>0$, 
we find that the pressure goes to zero for small temperatures since bosonic fluctuations are suppressed. With 
increasing temperature the pressure increases. In contrast to the results from LPA, the large-$N_c$ pressure
function reaches its maximal value already at $T=T_{\rm{c}}$. Thus, the large-$N_c$ results do not exhibit a 
maximum for $T>T_{\rm{c}}$. Comparing them with the results from LPA with a finite mass for the (pseudo-)
Goldstone boson, we find that the scalar fields still contribute significantly to the pressure of the system for 
$T/T_{\rm{c}} \lesssim 2$, even for comparatively large mass ratios $m_{\pi}/m_q\sim 1/2$. This is due to the fact 
that the thermal contributions to the masses of the scalar fields dominate for $T>T_{\text{max}}$. 
Therefore we conclude that in our studies the pressure below $T_{c}$ is mostly affected by the the presence of a 
finite quark mass. However, the maximum of the pressure is still clearly visible even though it is washed out 
compared to the chiral limit.

\begin{figure}[t]
\begin{center}
\includegraphics[scale=0.68,clip=]{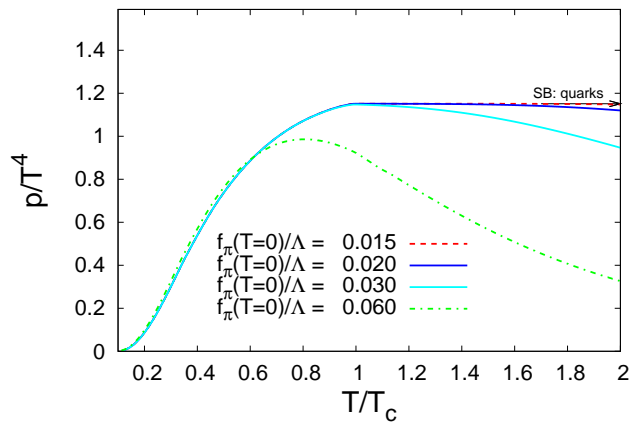}
\end{center}
\vspace*{-4mm}
\caption{(Color online)
Pressure as a function of $T/T_{\rm{c}}$ for various values of the UV cutoff $\Lambda$ as obtained from the RG flow
equations in a large-$N_c$ approximation. The parameters of the model for the various values of $\Lambda$ 
have been adjusted such that the IR physics at T=0 remain unchanged.
}\label{fig:cutoff}
\vspace*{-4mm}
\end{figure}

Finally we comment on a possible dependence of our results on the UV cutoff $\Lambda$. We illustrate such a 
dependence using our results for the large-$N_c$ approximation as an example, but we have checked for the other 
truncations as well that our results are not spoiled by the presence of the UV cutoff.  In Fig.~\ref{fig:cutoff}, we show 
the large-$N_c$ result for the pressure as a function of temperature for 
$f_{\pi}/\Lambda=0.060,\,0.030,\,0.020,\,0.015$. In the studied temperature range we find that the results for the 
pressure are strongly affected by the presence of the UV cutoff for $f_{\pi}/\Lambda \gtrsim 0.020$.
Note that we have chosen a 3$d$ regulator function which allows us to perform the Matsubara sum analytically.
Nonetheless the UV cutoff still enters into the thermal distribution functions in the flow 
equation~\eqref{eq:pot_flow} due to their dependence on $T/k$. In order to ensure that our results are not affected 
by the cutoff, we have chosen $f_{\pi}/\Lambda = 0.015$ in the present work. Note that in the large-$N_c$ 
approximation the cutoff dependence of the thermodynamic observables, e.~g. the pressure, can in principle 
be fixed by simply adding the missing contribution (at least in the chiral limit), see e.~g. Ref.~\cite{Braun:2003ii}:
\be
\frac{p_{\Lambda}}{T^4} = 
\frac{4N_c N_f}{6\pi ^{2}}\int _{\Lambda}^{\infty }dx \frac{x^{3}}{\E^x + 1}\,.
\ee
As soon as we allow for momentum dependent vertices and/or finite quark masses such a simple prescription can 
no longer be found.
\section{Conclusions and Outlook}\label{sec:conclusions}
We have discussed the truncation dependence of the thermodynamics of a simple model
consisting of strongly interacting relativistic bosons and chiral fermions. Our results indicate that
the derivative expansion is a good and valid approximation for a study of the thermal
properties of Yukawa-type theories, such as the pressure or the behavior of the order
parameter below the critical temperature. We find that the pressure exhibits a maximum
above the critical temperature when we include bosonic fluctuations. The origin of the maximum
is related to the fact that the thermal masses of the scalar fields are still small in a small window
above the critical temperature and therefore contribute significantly to the equation of state. 
For very high temperatures, the pressure then approaches the Stefan-Boltzmann limit of a gas of free 
massless fermions from above. The convergence to the Stefan-Boltzmann limit is faster the better the 
momentum dependence of the vertices is resolved.

On the other hand, we have also shown that the critical temperature shows a stronger dependence on 
a variation of the truncation. In order to clarify whether our results for $T_{\rm{c}}$ can be considered
as having almost converged, we eventually need to fully resolve the momentum dependence of the propagators 
and vertices as it has been successfully done 
in Refs.~\cite{Ellwanger:1995qf,Pawlowski:2003hq,Fischer:2004uk,Blaizot:2005xy,Blaizot:2005wd,Blaizot:2006vr,Diehl:2007xz,Benitez:2009xg}
at vanishing temperature. This is deferred to future work.

Whether the present findings, in particular the behavior of the pressure above $T_{\rm{c}}$, are
visible in the thermodynamic pressure of QCD can of course not be answered by the
model used in this study. First of all, the model is not confining and therefore the pressure
below $T_{\rm{c}}$ is overestimated. Within a functional RG approach, this can be resolved
by combining the findings of recent functional RG studies of QCD including gauge degrees of 
freedom~\cite{Gies:2002hq,Braun:2007bx,Braun:2008pi,Braun:2009gm}. 
This would allow us to compute the pressure from first principles in the sense that it would rely on only a 
single input parameter, namely the value of the strong coupling at the initial RG scale. Above the critical 
temperature, the pressure is mainly dominated by the presence of gluons. Still, the contribution from the 
mesons in the vicinity of the phase transition can be estimated from Eq.~\eqref{eq:max2flavor}
to yield a $10\%$ correction in QCD with two massless quark flavors. Therefore it might be 
possible to observe this effect in future lattice or functional RG studies of the thermodynamic pressure.

\acknowledgments
The author is very grateful to Holger Gies, Bertram Klein and Jan Martin Pawlowski for enlightening 
discussions and useful comments on the manuscript. This work was partly supported by the 
Natural Sciences and Engineering Research Council (NSERC) and by the National Research 
Council of Canada.
\appendix
\section{Decomposition of momentum tensors at finite temperature}\label{sec:decomposition}
Throughout this paper we encounter integrals of the following type
\be
{\mathcal I}(T)&=&T\sum_{n=-\infty}^{\infty}\int \frac{d^{d-1}}{(2\pi)^3}p_{\mu}p_{\nu}f(p_0^2, \vec{p}^2)\,,\\
\ee
where $T$ denotes the temperature, $p_0$ is a discrete variable depending on an integer $n$ and the function 
$f(.,.)$ simply represents a placeholder for a general function which is symmetric under $p_{\mu}\to -p_{\mu}$. 
Due to symmetry, the tensor structure can be simplified as follows:
\be
p_{\mu}p_{\nu}\rightarrow p_0^2 P^{\parallel}_{\mu\nu} + \frac{\vec{p}^{\,2}}{d-1}P^{\perp}_{\mu\nu}
\ee
where $n_{\mu}=(1,\vec{0})$ denotes the heat-bath velocity and the transversal ($\perp$) and the 
longitudinal ($\parallel$) projectors are defined as follows:
\be
P^{\parallel}_{\mu\nu}=n_{\mu}n_{\nu}\quad\text{and}\quad P^{\perp}_{\mu\nu}=\delta_{\mu\nu} - P_{\mu\nu}^{\parallel}\,.
\ee
\section{Threshold functions}\label{sec:thresholdfcts}
The regulator dependence of the flow equations is controlled by (dimensionless) threshold functions which arise 
from Feynman graphs, incorporating fermionic and/or bosonic fields. Let us first introduce the so-called 
dimensionless regulator-shape function $r_{\phi}(x)$ and $r_{\psi}(x)$ for bosonic and fermionic fields. These 
functions  are implicitly defined by the regulator function $R_i$ as follows:
\be
R_{\phi} (p_0,\{ p_i\}) = Z^{\perp}_{\phi} (p_0,\{ p_i\}) \vec{p}^{\,2} r_B (\vec{p}^{\,2}/k^2)
\ee
for the bosonic fields and
\be
R_{\psi} (p_0,\{ p_i\}) = Z^{\perp}_{\psi,k} (p_0,\{ p_i\}) \slash\!\!\!\vec{p}\,r_{\psi} (\vec{p}^{\,2}/k^2)
\ee
for the fermionic fields. 

In this work, we employ a 3$d$ optimized regulator-shape function~\cite{Braun:2003ii,Litim:2006ag,Blaizot:2006rj}:
\be
r_{\phi} (x)=\left(\frac{1}{x} - 1\right)\Theta (1-x)\label{eq:optshapefct_phi}
\ee
and
\be
r_{\psi}(x)=\left(\frac{1}{\sqrt{x}}-1\right)\Theta (1-x)\,.\label{eq:optshapefct_psi}
\ee
In the following, we use these regulator shape functions whenever we evaluate the integrals and sums in our 
general definitions of the threshold functions. Those threshold functions which are not defined below can be
found in Ref.~\cite{Braun:2008pi}.

In order to define the threshold functions, it is convenient to define dimensionless propagators for the scalar 
fields ($\phi$) and the fermions ($\psi$), respectively:
\be
\tilde{G} _{\phi} (x_0^{\phi},\omega)&=&\frac{1}{\hat{z}_{\phi} x_0^{\phi} + x(1+r_{\phi}) + \omega}\\
\text{and}\quad\tilde{G} _{\psi} (x_0^{\psi},\omega)&=&\frac{1}{\hat{z}_{\psi} ^2 x_0^{\psi} + x(1+r_{\psi})^2 + \omega}\,,
\ee
where $x_0^{\phi}=2\pi n \tilde{t}$ and $x_0^{\psi}=2\pi (n+1/2) \tilde{t}$. Note that
the Matsubara modes are dressed
with the ratio of the wave-function renormalizations longitudinal and transversal to the heat-bath, 
$\hat{z}_{\phi}=Z^{\parallel}_{\phi}/Z^{\perp}_{\phi}$ and $\hat{z}_{\psi}=Z^{\parallel}_{\psi}/Z^{\perp}_{\psi}$.
Moreover it is convenient to define the dimensionless propagators as obtained from the 
application of the optimized regulator-shape function defined in Eqs.~\eqref{eq:optshapefct_phi} 
and~\eqref{eq:optshapefct_psi}:
\be
{\mathcal G} _{\phi} (x_0^{\phi},\omega)&=&\frac{1}{1+ \hat{z}_{\phi} x_0^{\phi} + \omega}\\
\text{and}\quad {\mathcal G} _{\psi} (x_0^{\psi},\omega)&=&\frac{1}{1 + \hat{z}_{\psi} ^2 x_0^{\psi}  + \omega}\,.
\ee

For the wave-function renormalization of the scalar fields, we find:
\be
&&\hspace{-0.3cm}{\mathcal M}_{4,{\perp}} ^{(F)}(\tilde{t},\omega;\eta_{\psi})\nn\\
&&=(d\!-\!1)\tilde{t}\sum_{n=-\infty}^{\infty} \int _0 ^{\infty} \!dx x^{\frac{d-3}{2}}\tilde{\partial}_t
\Bigg\{ x (1\!+\! r_{\psi}) \tilde{G} _{\psi} (x_0^{\psi},\omega_{\psi})\!\times\nn\\
&&\times\Bigg[\frac{2x}{d-1} \left(\frac{d^2}{dx^2}(1+r_{\psi}) \tilde{G} _{\psi} (x_0^{\psi},\omega_{\psi})\right)\nn\\
&&\qquad + \frac{d+1}{d-1} \left(\frac{d}{dx}(1+r_{\psi}) \tilde{G} _{\psi} (x_0^{\psi},\omega_{\psi})\right)\Bigg]\nn\\
&& +\, \hat{z}_{\psi}^2 x_0^{\psi}  \tilde{G} _{\psi} (x_0^{\psi},\omega_{\psi}) 
\Bigg[
\frac{2x}{d-1} \left( \frac{d^2}{dx^2} \tilde{G} _{\psi} (x_0^{\psi},\omega_{\psi}) \right)\nn\\
&& \qquad\qquad\qquad\qquad\qquad\; +  \left( \frac{d}{dx} \tilde{G} _{\psi} (x_0^{\psi},\omega_{\psi}) \right)
\Bigg]\Bigg\}.
\ee
Employing the optimized 3$d$ regulator function specified above, we obtain
\be
&&\hspace{-0.3cm}{\mathcal M}_{4,{\perp}} ^{(F)}(\tilde{t},\omega)\nn\\
&&=4\tilde{t}\sum_{n=-\infty}^{\infty} 
\Bigg\{ (1+ z_{\psi}^2 x_0^{\psi})\left( {\mathcal G} _{\psi} (x_0^{\psi},\omega_{\psi}) \right)^4\nn\\
&& +\frac{1}{d\!-\!3} \left( {\mathcal G} _{\psi} (x_0^{\psi},\omega_{\psi}) \right)^3  
- \left(\frac{1}{2d\!-\!6}\!+\!\frac{1}{4}\right)\left( {\mathcal G} _{\psi} (x_0^{\psi},\omega_{\psi}) \right)^2 \nn\\
&&\qquad -\frac{\eta_{\psi}^{\perp}}{d\!-\!3} \left( {\mathcal G} _{\psi} (x_0^{\psi},\omega_{\psi}) \right)^3
 + \frac{\eta_{\psi}^{\perp}}{2d\!-\!6}\left( {\mathcal G} _{\psi} (x_0^{\psi},\omega_{\psi}) \right)^2\Bigg\}\,.\nn\\
\ee
 The threshold function for the corresponding longitudinal contribution reads:
 \be
&&\hspace{-0.3cm}{\mathcal M}_{4,{\parallel}} ^{(F)}(\tilde{t},\omega)\nn\\
&& =\tilde{t}\sum_{n=-\infty}^{\infty} \int _0 ^{\infty} \!dx x^{\frac{d-3}{2}}\tilde{\partial}_t
\Bigg\{ x (1\!+\! r_{\psi}) \tilde{G} _{\psi} (x_0^{\psi},\omega_{\psi})\!\times\nn\\
&&\times\Bigg[2 x_0^{\psi} \left(\frac{d^2}{(dx_0^{\psi})^2}(1+r_{\psi}) \tilde{G} _{\psi} (x_0^{\psi},\omega_{\psi})\right)\nn\\
&&\qquad + \left(\frac{d}{dx_0^{\psi}}(1+r_{\psi}) \tilde{G} _{\psi} (x_0^{\psi},\omega_{\psi})\right)\Bigg]\nn\\
&& +\, \hat{z}_{\psi}^2 x_0^{\psi}  \tilde{G} _{\psi} (x_0^{\psi},\omega_{\psi}) 
\Bigg[
2 \left( \frac{d^2}{(dx_0^{\psi})^2} \tilde{G} _{\psi} (x_0^{\psi},\omega_{\psi}) \right)\nn\\
&& \qquad\qquad\qquad\quad\; +  3\left( \frac{d}{dx_0^{\psi}} \tilde{G} _{\psi} (x_0^{\psi},\omega_{\psi}) \right)
\Bigg]\Bigg\}\nn
\ee
\be
&&=z_{\psi}^2\tilde{t} \sum_{n=-\infty}^{\infty}\frac{2}{d-1}\left(1-\frac{\eta_{\psi}^{\perp}}{d}\right)\Bigg\{
-2 \left( {\mathcal G} _{\psi} (x_0^{\psi},\omega_{\psi}) \right)^3 \nn\\
&& \qquad\qquad\qquad +(26\hat{z}_{\psi}^2 x_0 ^{\psi} +6)  \left( {\mathcal G} _{\psi} (x_0^{\psi},\omega_{\psi}) \right)^4\nn\\
&&\qquad
-32 \hat{z}_{\psi}^2 x_0^{\psi} (1 +  \hat{z}_{\psi}^2 x_0) \left( {\mathcal G} _{\psi} (x_0^{\psi},\omega_{\psi}) \right)^5\Bigg\}\,.
\ee
Apart from these contributions, the scalar wave-function renormalization also gets a contribution from a fermion loop 
which is proportional to the vacuum expectation value of the scalar field:
\be
&&\hspace{-0.3cm}{\mathcal M}_{2,{\parallel}} ^{(F)}(\tilde{t},\omega_{\psi})
=\frac{1}{4}\hat{z}_{\psi}^2\tilde{t}\sum_{n=-\infty}^{\infty}\int _0 ^{\infty} \!dx x^{\frac{d-3}{2}}\tilde{\partial}_t
\tilde{G} _{\psi} (x_0^{\psi},\omega_{\psi})\times\nn\\
&&\qquad\qquad\qquad\qquad\qquad\qquad\times\Bigg\{
\left(\frac{d}{d x_0^{\psi}} \tilde{G} _{\psi} (x_0^{\psi},\omega_{\psi})\right)\nn\\
&& \qquad\qquad\quad
+ 2\hat{z}_{\psi}^2 x_0^{\psi} 
\left(\frac{d^2}{(d x_0^{\psi})^2} \tilde{G} _{\psi} (x_0^{\psi},\omega_{\psi})\right)
\Bigg\}\nn\\
&&
=\frac{1}{d-1}\left(1-\frac{\eta_{\psi}^{\perp}}{d}\right)\hat{z}_{\psi}^2\tilde{t}\sum_{n=-\infty}^{\infty}
\left({\mathcal G} _{\psi} (x_0^{\psi},\omega_{\psi})\right)^4\times\nn\\
&&\qquad\qquad\qquad\quad\times\left\{
3- 16 \hat{z}_{\psi}^2 x_0^{\psi}{\mathcal G} _{\psi} (x_0^{\psi},\omega_{\psi})
\right\}
\,.
\ee
Finally, there is a contribution to the scalar wave-function renormalization from a pion-sigma loop 
which is proportional to the four-boson coupling:
\be
&&\hspace{-0.3cm}{\mathcal M}_{2,2,{\parallel}} ^{(B)}(\tilde{t},\omega_{\pi},\omega_{\sigma})\nn\\
&&
\hspace{-0.3cm}=\frac{\hat{z}_{\phi}\tilde{t}}{8}\!\sum_{n=-\infty}^{\infty}\int _0 ^{\infty} \!dx x^{\frac{d-3}{2}}\tilde{\partial}_t
{\mathcal G} _{\phi} (x_0^{\phi},\omega_{\phi,1})\left( {\mathcal G} _{\pi} (x_0^{\phi},\omega_{\sigma}) \right)^2\!\times\nn\\
&&\qquad\times\left\{
8\hat{z}_{\phi}x_0^{\phi} {\mathcal G} _{\phi} (x_0^{\phi},\omega_{\sigma}) -2 
\right\} + ( \omega_{\pi} \leftrightarrow \omega_{\sigma})\nn\\
&&\hspace{-0.3cm}=\left(1\!-\!\frac{\eta_{\phi}^{\perp}}{d\!+\!1}\right)\! \frac{\hat{z}_{\phi}  \tilde{t}}{(d\!-\!1)}
\sum_{n=-\infty}^{\infty} \Bigg\{ {\mathcal G} _{\phi} (x_0^{\phi},\omega_{\pi})\times\nn\\
&&\qquad\quad\times\left( {\mathcal G} _{\phi} (x_0^{\phi},\omega_{\sigma})\right)^2
\left[ \,2 {\mathcal G} _{\phi} (x_0^{\phi},\omega_{\sigma}) + {\mathcal G} _{\phi} (x_0^{\phi},\omega_{\pi}) \right]\nn\\
&& - 4 \hat{z}_{\phi} x_0^{\phi}{\mathcal G} _{\phi} (x_0^{\phi},\omega_{\pi})\!\!\left( {\mathcal G} _{\phi} (x_0^{\phi},\omega_{\sigma})\right)^3
\big[ {\mathcal G} _{\phi} (x_0^{\phi},\omega_{\pi}) \nn\\
&&\qquad\qquad \qquad\;\; + 3\,{\mathcal G} _{\phi} (x_0^{\phi},\omega_{\sigma})\big]\Bigg\} + ( \omega_{\pi} \leftrightarrow \omega_{\sigma})\,.
\ee
For the fermionic wave-function renormalization longitudinal to the heat-bath, we find
\be
&&\hspace{-0.3cm}{\mathcal M}^{(FB)}_{1,2,{\parallel}} (\tilde{t},\omega_{\psi},\omega_\phi)\nn\\
&& =\frac{\tilde{t}}{2} \sum_{n=-\infty}^{\infty} \int _0 ^{\infty} \!dx x^{\frac{d-3}{2}}\tilde{\partial}_t
\frac{\hat{z}_{\psi} x_0^{\psi}}{(1+r_{\psi})}\Bigg\{ - \tilde{G} _{\psi} (x_0^{\psi},\omega_{\psi})\times\nn\\
&& \qquad\qquad\qquad\qquad\times\left(\frac{d}{d x_0^{\psi}}\tilde{G} _{\phi} (x_0^{\psi},\omega_{\phi})\right)
\Bigg\}\nn
\ee
\be
&& = \frac{2}{d-1}\hat{z}_{\psi}\hat{z}_{\phi}\tilde{t} \sum_{n=-\infty}^{\infty} x_0^{\psi}
{\mathcal G} _{\psi} (x_0^{\psi},\omega_{\psi})\left({\mathcal G} _{\phi} (x_0^{\psi},\omega_{\phi})\right)^2\times\nn\\
&&\hspace{-0.3cm}\times\Bigg\{
\left(1\!-\!\frac{\eta_{\psi}^{\perp}}{d}\right) {\mathcal G} _{\psi} (x_0^{\psi},\omega_{\psi})
+ 2 \left(1\!-\! \frac{\eta_{\phi}^{\perp}}{d\!+\!1}\right) {\mathcal G} _{\phi} (x_0^{\psi},\omega_{\phi})
\Bigg\}\,.\nn\\
\ee
Note that the Matsubara sums in the threshold functions can be performed analytically.  Since
the resulting expressions are rather complicated and of no interest for the discussions in the
present paper, we do not show them here. However, for the numerical evaluation of the RG flows
we have made use of the fact that the Matsubara sum in these functions can
be performed analytically.

\bibliography{references}

\end{document}